\documentclass{article}

\newcommand{\iec}{\mbox{i.\,e.\,}}

\newcommand{\egc}{\mbox{e.\,g.\,}}

\newcommand{\dr}[1]{\ensuremath{\mathrm{d} #1\,}}
\newcommand{\mc}[1]{\ensuremath{\mathcal{#1}}}

\newcommand{\ddt}{\ensuremath{\frac{\dr{}}{\dr{t}}}}

\newcommand{\ket}[1]{\ensuremath{\left|  #1 \right\rangle}}
\newcommand{\bra}[1]{\ensuremath{\left\langle #1 \right|}}
\newcommand{\bk}[2]{\ensuremath{\left\langle #1 | #2 \right\rangle}}
\newcommand{\proj}[2]{\ensuremath{\ket{#1} \bra{#2}}}
\newcommand{\tpk}[2]{\ensuremath{\ket{#1}\!\otimes\!\ket{#2}}}
\newcommand{\tpkthree}[3]{\ensuremath{\ket{#1}\!\otimes\!\ket{#2}\!\otimes \! \ket{#3}}}

\newcommand{\matel}[3]{\ensuremath{\bra{#1} #2 \ket{#3}}}
 
\newcommand{\op}[1]{\ensuremath{\widehat{\textsf{\ensuremath{#1}}}}}

\newcommand{\id}{\op{\mathsf{1}}}

\newcommand{\comm}[2]{\ensuremath{\left[ #1 , #2 \right]}}

\newcommand{\be}{\begin{equation}}
\newcommand{\ee}{\end{equation}}
\newcommand{\e}[1]{\mathrm{e}^{#1}}

\usepackage{chicago}
\begin{document}

\title{What is orthodox quantum mechanics?}
\author{David Wallace}
\maketitle

\begin{abstract}
What is called ``orthodox'' quantum mechanics, as presented in standard foundational discussions, relies on two substantive assumptions --- the projection postulate and the eigenvalue-eigenvector link --- that do not in fact play any part in practical applications of quantum mechanics. I argue for this conclusion on a number of grounds, but primarily on the grounds that the projection postulate fails correctly to account for repeated, continuous and unsharp measurements (all of which are standard in contemporary physics) and that the eigenvalue-eigenvector link implies that virtually all interesting properties are maximally indefinite pretty much always. I present an alternative way of conceptualising quantum mechanics that does a better job of representing quantum mechanics as it is actually used, and in particular that eliminates use of either the projection postulate or the eigenvalue-eigenvector link, and I reformulate the measurement problem within this new presentation of orthodoxy.
\end{abstract}

\section{Introduction: the orthodox view of orthodoxy}

``Orthodox'' or ``standard'' quantum mechanics, as typically presented in textbook philosophy-of-physics discussions\footnote{See, \egc, \citeN{albertqmbook}, \citeN{barrettbook}, \citeN{bubbook}, \citeN[ch.5--6]{penroseenm}}, consists of these components:
\begin{description}
\item[The structural core:] This has three parts:
\begin{enumerate}
\item \textbf{States:} The possible states of a quantum system are represented by normalised vectors in some complex Hilbert space.
\item \textbf{Observables:} To any physical quantity used to describe the system (often called an `observable') is associated a self-adjoint operator on that same Hilbert space.
\item \textbf{Dynamics:} The state of a quantum system evolves over time according to the \emph{Schr\"{o}dinger equation}:
\be 
\ddt \ket{\psi(t)} = -\frac{i}{\hbar}\op{H}\ket{\psi(t)}
\ee
where $\op{H}$ is the self-adjoint operator corresponding to the system's energy.
\end{enumerate}
(This can be generalised in certain respects, in particular by allowing quantum states to be mixed rather than pure, and in fact I think this generalisation does a better job of capturing real-world quantum mechanics than the pure-state version (cf \citeNP{wallaceinfdyn}) but for simplicity I use the pure-state version in this paper.)

\item[The Born (probability) rule]: Suppose some quantity $O$ has associated operator $\op{O}$, which can be written 
\be \label{specrel}
\op{O}=\sum_i o_i \op{\Pi}(i)
\ee
where the $o_i$ are the distinct eigenvalues of the operator and $\op{\Pi}(i)$ projects onto the subspace of states with eigenvalue $o_i$. (Recall that any self-adjoint operator can be so written --- this is the `spectral resolution' of the operator.) Then if $O$ is measured on a quantum system with state $\ket{\psi}$, then:
\begin{enumerate}
\item The only possible outcomes of the measurement are the eigenvalues $o_i$ of the operator;
\item The probability of the measurement giving result $o_i$ is
\be 
\Pr(O=o_i)=\matel{\psi}{\op{P}(i)}{\psi}.
\ee
\end{enumerate}
\item[The projection postulate (aka the collapse law):] Suppose some quantity $O$, as above, is measured on a quantum system in state \ket{\psi}. Then the measurement induces a stochastic transition on the state, so that:
\begin{enumerate}
\item Immediately after the measurement, the system is in one of the states
\be 
\ket{\psi_i} = \frac{\op{\Pi}(i)\ket{\psi}}{\|\op{\Pi}(i)\ket{\psi}\|}.
\ee
\item The probability that the system transitions into state $\ket{\psi_i}$ is given by
\be \label{collapsetransprob}
\Pr(\ket{\psi}\rightarrow \ket{\psi_i})= \matel{\psi}{\op{P}(i)}{\psi}.
\ee
\end{enumerate}
(The projection law thus restricts the generality of the Schr\"{o}dinger equation: systems evolve under it \emph{only when a measurement is not taking place}.)
\item[The eigenvector-eigenvalue link (E-E link):] Given an quantity $O$ as above:
\begin{enumerate}
\item A system in state \ket{\psi} possesses a definite value of $O$ if and only if \ket{\psi} is an eigenstate of \op{O}, $\op{O}\ket{\psi}=o_i\ket{\psi}$.
\item In this case, the definite value is the associated eigenvalue $o_i$.
\end{enumerate}
\end{description}
Given one additional assumption --- that if a measurement of $O$ returns value $o_i$, the measured system actually has value $o_i$ of $O$ --- the Born rule can be derived from the projection postulate and the eigenvalue-eigenvector link. For if $O$ is measured on a system in state $\ket{\psi}$, by the projection postulate it will transition into an eigenstate of \op{O}, with the probability of transitioning given by (\ref{collapsetransprob}); after the collapse, it will have a definite value of $O$ by the eigenvalue-eigenvector link; if measurement simply reports that definite value, the Born rule follows.

In any case, it is standard in foundations of quantum mechanics to treat both the projection postulate and the E-E link as core components of orthodox QM. Interpretations of QM like Everett's and Bohm's, for instance, are specifically described as `no-collapse' interpretations in view of the fact that they drop the collapse law from the postulates of QM; discussions of the ontology of the GRW collapse theory \citeN{albertloewer1996} talk of the need to abandon the E-E link; attempts at interpretation-neutral discussions of the ontology of QM (\egc, \citeNP{skow2010,darby2010,bokulich2014,wolff2015,wilsonjowett}) typically take the E-E link as a starting point.

Furthermore, typical statements of the quantum measurement problem typically take the E-E link, and/or the projection postulate, as central. The measurement problem is the problem of \emph{macroscopic indefiniteness}, of quantum states that describe macroscopic systems in states that are indefinite with regard to ordinary properties such as the location of pointers or the heartbeats of cats. Or --- if macroscopic indefiniteness is to be removed via the projection postulate --- it is the problem of \emph{dynamical ill-definedness}, of the lack of any well-defined recipe as to when collapse occurs (it is easy to show that collapse cannot be a consequence or special case of the Schr\"{o}dinger equation applied to a complex measuring system).

The purpose of this paper, by contrast, is to argue that orthodox quantum mechanics in fact consists only of the structural core and the Born rule. The projection postulate, and the eigenvector-eigenvalue link, are at best parts of a proposed interpretation of QM that goes beyond orthodoxy, at worst unmotivated distractions. As such, in formulating (as opposed to solving) the quantum measurement problem, we should begin with just the structural core and the Born rule. We might  introduce one or both as part of a \emph{solution} to the measurement problem, but we confuse the dialectic by taking them as initial common ground.

To be clear what I mean: I will not (here) argue that the \emph{best} or \emph{right} way to interpret QM, or to solve the quantum measurement problem, involves abandoning collapse and/or the E-E link. I will argue that QM as actually practiced by physicists --- and what does ``orthodox'' QM mean, if not that? --- already proceeds without either.

In section \ref{vscollapse-indirect} I provide some evidence that physicists in practice do not seem to make use of a collapse rule. I strengthen this in sections \ref{vscollapse-repeated}--\ref{vscollapse-continuous} by arguing that the collapse rule is incapable of handling two standard kinds of experimental setup: those involving repeated measurements, and those involving continuous observation. In section \ref{vsEE-position} I point out that the rapid spreading of wavepackets under the Schr\"{o}dinger equation means that the E-E link makes the ridiculous claim that essentially any system, including macroscopic systems, is maximally indefinite in position, and hence that the E-E link does not have the resources to say when systems are actually localised; in section \ref{vsEE-hegerfeldt} I deploy a result of Hegerfeldt to show that this generalises to pretty much any observable.

For the remainder of the paper, I explore what `orthodox quantum mechanics' is, shorn of the E-E link and the projection postulate. (This part of the paper draws on some parts of \citeN{wallacecosmology}, albeit deployed in a rather different context). I consider (in section \ref{labview}) a view which treats preparation and measurement as primitive, but ultimately reject it (in section \ref{labview-limitation}) on the grounds both that it too struggles with continuous and repeated measurements, and that it cannot handle applications of QM where results from QM are integrated into larger pieces of historical science. With this as a starting point, I finally suggest (section \ref{decoherentview}) that orthodoxy should be understood as an inchoate attitude to the quantum state, where its dynamics are always unitary but where it is interpreted either as physically representational or as probabilistic, according to context. In the concluding section I reflect on the right formulation of the measurement problem given this conception of what `orthodox QM' actually is.

\section{Against collapse: indirect evidence}\label{vscollapse-indirect}

The projection postulate appears in \citeN{Dirac1930} and \citeN{vonneumann}, the first two codifications of the axioms of QM. It continues to be widespread, though not universal, in first courses on QM to this day: an unscientific perusal of my shelf reveals that collapse is included in about half of the  books there that present QM from scratch. (The Born rule, of course, appears in all of them.) But for all this (I will argue) it plays no real role in applications of quantum mechanics in physics. It is rather hard to prove a negative, but here I give some suggestive reasons to think that physical practice abjures collapse.

Firstly, collapse is conspicuously absent from \emph{second} courses in QM,  and in particular in courses on relativistic QM. This ought to strike a student as peculiar (it certainly struck the author, as an undergraduate, as peculiar): collapse, as formally defined in QM, is a \emph{global} phenomenon, applying to the whole quantum state and so affecting, simultaneously, systems spatially far from one another. In relativity, this notion of simultaneity is frame-dependent (or simply meaningless, depending how you think about conventionality of simultaneity, but in any case problematic). One would expect, if collapse is really part of orthodox QM, that the first chapter of any relativistic QM textbook would start with a careful discussion of exactly how the collapse postulate is to be applied in the relativistic context. I have not once seen any such textbook so much as consider the question.

Again, to be clear: the point is not that collapse is \emph{unsatisfactory} in the relativistic regime. Of course it is; the tension between relativity and QM has been known at least since the EPR paper \cite{epr}. But relativistic QM textbooks contain, not an unsatisfactory collapse rule, but no collapse rule at all. One concludes that the theory must be applicable without any mention of collapse. And indeed it is: the name of the game in relativistic QM is to calculate probablility distributions over physical quantities --- most often, over the various energies, momenta and particle numbers of the decay products of some scattering experiment --- and for this, only the Born rule is required; collapse plays no part.

Secondly, the theoretical physics community has been worrying for forty years about the so-called ``black hole information loss paradox'' originally identified by \citeN{hawking1976}. (See, \egc, \citeN{pageinformationlossreview,belotetalinformationloss} and references therein, though the debate continues in lively fashion to this day.) At its heart, the paradox is simply that black hole decay is \emph{non-unitary} and as such can't be described within the Schr\"{o}dinger-equation framework. But state-vector collapse is also non-unitary! So if the collapse law is part of orthodox QM, quantum-mechanical dynamics were never unitary in the first place: they were an alternating series of unitary and non-unitary processes. So why be so desparate to preserve unitary in the exotic regime of black hole decay, when it is ubiquitous in far more mundane cases? One has the clear impression that (at least this part of) the theoretical physics community does \emph{not} in fact think that dynamics is non-unitary in any other contexts in physics, rendering black hole decay uniquely problematic. Tempting though it might be for this advocate of the Everett interpretation to claim that the community has adopted the many-worlds theory \emph{en masse}, a more mundane account is simply that (what they regard as) \emph{orthodox} QM does not include the collapse postulate.\footnote{Of course, plenty of people working on black hole decay \emph{are} fairly explicit advocates of the Everett interpretation, and I have argued elsewhere that quantum cosmology generally is \emph{tacitly} committed to the Everett interpretation, but it's clear that the majority of the community embrace Mermin's ``shut up and calculate'' approach~\cite{merminfeynman}.} 

Thirdly, modern quantum field theory largely abandons Hamiltonian methods in favour of the path-integral approach. But in that approach it is not even clear how collapse is to be defined (and, again, textbook presentations never seem to mention the issue), and yet the theory still seems to produce empirically successful predictions.

Finally, and as an admittedly crude indicator, searching the archives of \emph{Physical Review} for “projection postulate”, “wave-function collapse” and the like turns up only a few hundred references, nearly all of which turn out to be (a) foundational discussions, (b) discussions of proposed alternatives to quantum theory, or (c) theoretical quantum-computation discussions. (For comparison, searches for terms like “state vector” or “Hilbert space” or “Schrodinger equation” typically turn up several tens of thousands of references.)

\section{Against collapse: inadequacy for repeated measurements}\label{vscollapse-repeated}

The case of repeated measurements --- when some quantity is measured on a quantum system and then, a short while later, measured again --- has actually been used, since Dirac, as an argument \emph{for} state-vector collapse. The argument goes like this: repeated measurements must give identical results; so if a measurement of  $O$ gives outcome $o_i$, then a subsequent measurement of $O$ immediately afterwards must also give outcome $o_i$, The only way this is compatible with the Born rule is if the state of the system immediately before this second measurement is an eigenstate of $\op{O}$ with eigenvalue $o_i$ --- so to get repeated measurements right, wavefunction collapse is a requirement.

\ldots which would be all very well, if repeated measurements \emph{did} give identical results. But:
\begin{itemize}
\item photon detectors typically absorb photons: immediately after a measurement on a photon, the photon no longer exists;
\item The Stern-Gerlach apparatus detects an atom's spin by slamming it very hard into a screen; this process is in no way guaranteed to preserve that atom's spin.
\item More generally, measuring something by slamming it very hard into something else is probably the single most commonly used tool in the experimental physicist's toolbox.
\end{itemize}
In fact, `non-disturbing' measurements, in which repeated measurements indeed give the same results, are decidedly uncommon in quantum mechanics and require some skill to set up (see \citeN{homewhitakerzeno} for discussion). So a collapse rule explicitly designed to ensure that repeated measurements give the same results is in flat conflict with a lot of observed physics.

By contrast, quantum mechanics \emph{without} collapse has no trouble with repeated measurements --- non-disturbing or otherwise. The familiar trick, following von Neumann's original prescription, is to include the measurement device in the physical analysis. Suppose for simplicity that \op{O} is non degenerate,
\be 
\op{O}=\sum_i o_i \proj{o_i}{o_i},
\ee
and suppose that the measurement device has some observable $M$ corresponding to the possible measurement outcomes. In von Neumann's original version, $M$ is the position of the centre of mass of some pointer; here for convenience I take $\op{M}$ too as being discrete and nondegenerate,
\be
\op{M}=\sum_i m_i \proj{m_i}{m_i}.
\ee
Then the measurement interaction is assumed to have form
\be 
\tpk{o_i}{m_0}\rightarrow \tpk{\varphi_i}{m_i}.
\ee
Applying this measurement process to a system initially in state 
\be 
\ket{\psi}=\sum_i \lambda_i \ket{o_i}
\ee
and a measurement device initially in state $\ket{m_0}$ gives the outcome
\be 
\tpk{\psi}{m_0}\rightarrow \sum_i \lambda_i \tpk{\varphi_i}{m_i}.
\ee
Applying the Born rule to a measurement \emph{of M} now tells us that the probability of getting $m_i$ is $|\lambda_i|^2$ --- exactly what the Born rule requires for a measurement of $O$ on the original system, and it is for exactly this reason that this process indeed qualifies as a measurement.

There is no requirement here that $\ket{\varphi_i}=\ket{o_i}$ or even that the distinct $\ket{\varphi_i}$ are orthogonal --- indeed, the measurement process could perfectly well dump the measured system in some fixed post-measurement state $\ket{\varphi_0}$ (as in the case of photon absorption) in which case the measurement process is
\be 
\tpk{\psi}{m_0}\rightarrow \ket{\varphi_0} \otimes \sum_i \lambda_i \ket{m_i}.
\ee
The `non-disturbing' measurements are then the ones where indeed $\ket{\varphi_i}=\ket{o_i}$. In these cases, but only these, if we bring in a second copy of the measurement device and repeat the measurement interaction, we get 
\be 
\tpkthree{\psi}{m_0}{m_0}\rightarrow \left(\sum_i \lambda_i \tpk{o_i}{m_i}\right)\otimes\ket{m_0} \rightarrow \sum_i \lambda_i \tpkthree{o_i}{m_i}{m_i}.
\ee
Applying the Born rule in this case to a joint measurement of $M\times M$, we find that indeed, when measurements are non-disturbing the probability is 100\% that two successive measurements give the same result.

If there is a lesson to learn from repeated measurements it is that the Born rule, by itself, does not define \emph{transition} probabilities, but only probabilities at an instant (an issue I return to in section \ref{decoherentview}). But ``wave functions collapse on measurement'' does not solve this problem satisfactorily, and indeed gives flatly incorrect results.

\section{Against collapse: inadequacy for continuous measurement}\label{vscollapse-continuous}

Continuous measurements --- where a system is constantly observed to see if, or how quickly, it undergoes some change --- are commonplace in physics. For instance, radioactive decay measurements --- where a Geiger counter is placed near some radioactive substance, and the rate of decay is recorded --- are among the most straightforward demonstrations of quantum mechanics' probabilistic nature. Yet they fit strikingly badly into the wavefunction-collapse framework. 

It is not that the \emph{physics} of (for instance) radioactive decay is problematic, at least phenomenologically. (Actually calculating decay rates \emph{ab initio} is another matter: the nucleus is a complex and strongly bound system, and hard to treat analytically.) The idea is that --- if the decay rate is $1/\tau$ --- then an undecayed particle, in state $\ket{\mbox{undecayed}}$, evolves over some short time interval $\delta t$ like 
\be 
\ket{\mbox{undecayed}} \rightarrow \left(1- \frac{\delta t^2}{2\tau} \right)\ket{\mbox{undecayed}} + \sqrt{\frac{1}{\tau}}\delta t \ket{\mbox{decay products}}.
\ee
Meanwhile, the decay-product state's own evolution over time, which can be represented as
\be 
\ket{\mbox{decay products}}\rightarrow \op{U}(t)\ket{\mbox{decay products}}\equiv \ket{\mbox{decay products;}t}
\ee
explores a very large region of Hilbert space and, in particular, satisfies
\be
\bk{\mbox{decay products}}{\mbox{decay products;}t}\simeq 0 
\ee
for $t_0<t<T$, where $T$ is the (extremely large) Poincar\'{e} recurrence time for the system and $t_0 \ll \tau$ (\iec, the rate at which the radioactive products evolve away from their original state is much quicker than the particle's decay rate). Under these assumptions we can deduce
\be 
\op{U}(t)\ket{\mbox{undecayed}}=\e{-t/2\tau}\ket{\mbox{undecayed}} + \int_0^t \dr{\xi} \frac{\e{-\xi/2\tau}}{\sqrt{\tau}}
\ket{\mbox{decay products;}(t-\xi)}
\ee
at least for $T \gg t \gg \tau$.

Applying the Born rule to this system gives exactly the results we would expect: at time $t$, the probability of the system being undecayed is 
$|\e{-t/2\tau}|^2=\e{-t/\tau}$. And no assumption of wavefunction collapse is required to derive this probability. But suppose we make that assumption anyway: when, in that case, is the wavefunction supposed to collapse? 

One possibility would be to model the continuous process of measurement by a frequent but discrete series --- applying the projection postulate every $\Delta$ seconds --- and then taking $\Delta \rightarrow 0$. As long as $\Delta$ is long enough  --- technically speaking: as long as 
$\Delta \gg t_0$ --- this iterated collapse will leave the probabilities unaffected. But it is the content of the quantum Zeno paradox \cite{misrasudarshan} that as $\Delta \rightarrow 0$, the evolution of the system is entirely halted: in this limit, the state of the particle remains $\ket{\mbox{undecayed}}$ forever.

Misra and Sudarshan did assume (at least for the purposes of their paper) that observation required collapse, and so that continuous observation required continuous collapse; hence ``paradox''. But it is the collapse postulate, not anything about continuous observation \emph{per se}, that delivers this impossible result. Modelling of measurement as a physical process, as per the previous section, reveals (cf \citeNP{homewhitakerzeno}) that:
\begin{itemize}
\item The Zeno `paradox' is a real (and empirically confirmed) physical effect: if a discrete measurement process is carried out repeatedly (and, crucially, if the time taken to carry out every individual measurement is short compared to the timescales on which the measured system evolves), then the rate of evolution of the system really is reduced by the measurements, and tends to zero as the frequency of repetition tends to infinity.
\item A continuous observation can also be modelled as a physical process, and in this case the relevant variable is the response speed of the measurement device compared to the timescale on which the measured system evolves. Again, when the former is much faster than the latter, evolution is heavily suppressed. But an observation can be `continuous' even while its response time is relatively slow: in the case of a Geiger counter, the response time is so slow compared to the evolution timescales of decay that Zeno slowing is negligible. (The relevant system timescale is not the decay rate, but the evolution time of the decay products, i.e. $t_0$.)
\end{itemize}
(This is not to say that the Zeno effect is \emph{entirely} non-paradoxical, even when understood without the distorting reference to collapse. Paradoxical (though non-contradictory) consequences arise when the measurement process involves energy exchange between measurement device and system only when the system is in a state distinct from its original state, so that the presence of the measurement device appears to halt the system's evolution even though the two are not interacting. This is related to the phenomenon of interaction-free measurement, as seen in the Elitzur-Vaidman bomb problem~\citeN{elitzurvaidman}. For further discussion see Home and Whitaker, \emph{ibid}, or (for an unapologetically pro-Everettian perspective!) \citeN[pp.390--3]{wallacebook}.)

\section{Against collapse: inadequacy for unsharp measurement}

The view that measurements are represented by collections of mutually orthogonal projectors is now thirty years out of date. Quantum measurement theory now regards the ``projection-valued measurements'' (PVMs) that can be so represented merely as a special case of a more general framework: ``positive-operator-valued measurements'' (POVMs).\footnote{For more detail on the physics of this section, see, \egc, \citeN{buschmeasurement}.} In the POVM framework, measurements are represented by collections $\{\op{M}_i\}$ of self-adjoint operators that (i) are positive (that is, have no negative eigenvalues, or equivalently, satisfy $\matel{\psi}{\op{M}_i}{\psi}\geq 0$ for any state $\ket{\psi}$); (ii) sum to unity, $\sum_i \op{M}_i = \id$.

For instance, consider measuring a particle's phase-space position: that is, consider simultaneously measuring its position and its momentum. Within the PVM framework, this is impossible: position and momentum do not commute. But in modern measurement theory, this simply tells us that we cannot make a simultaneous \emph{sharp} measurement of position and momentum. We can measure both provided we are prepared to accept a little noise in the measurement process, and for macroscopically large systems the noise can be very small indeed --- which is reassuring, since manifestly we \emph{do} simultaneously measure the position and momentum of macroscopic bodies.

A phase-space POVM (in, for simplicity, one spatial dimension) can be defined by starting with some state $\ket{\varphi}$ that is a wavepacket approximately localised around position and momentum zero (say, a Gaussian), so that
\be 
\ket{\varphi(p,q)}=\exp{(-i\op{X}p)}\exp{(+i \op{P}q)}\ket{\varphi}
\ee
is the same state translated so as to be localised around position $q$ and momentum $p$. Then the family of operators
\be 
\op{M}_{p,q}=\frac{1}{2\pi}\proj{\varphi(p,q)}{\varphi(p,q)}
\ee
is a POVM and can be used to represent the unsharp phase-space measurement. It will give probability distributions over position and momenta separately which are smearings-out of the sharp results obtained from the Born rule, with the level of smearing depending on the width of the wavepacket in position and momentum space and becoming negligible in both cases for macroscopically large systems.

Similarly, the POVM framework can handle fuzzy measurements of a single quantity, as might be appropriate when the measurement device is imperfect. Given an observable $O$ corresponding to an operator $\op{O}$ with spectral resolution (\ref{specrel}), suppose that $f^1,\ldots f^N$ are $N$ functions from the spectrum of $\op{O}$ to the nonnegative reals, satisfying
\be
\sum_{k=1}^N f^k(o_i)=1
\ee
for all $i$. Then the family of operators $f_k(\op{O})$ is a POVM. If $f_k(o_i)=\delta^k_i$, this just reduces to a sharp measurement of $O$, but more general measurements of $O$ can be represented by more general choices of the $f^k$.

The POVM generalisation of traditional measurement theory is by now routine, and mathematically speaking is a straightforward generalisation of the Born rule. But it has no associated collapse law, and so it is opaque how to apply collapse in POVM contexts. In addition, POVMs are not associated with the spectral decompositions of the operators representing physical quantities, so to deduce what POVM is being applied, we need to model the measurement process as a unitary interaction with the measurement device, and then in due course apply the Born rule with respect to a macroscopic quantity pertinent to the measurement device.

The lessons of continuous, repeated and unsharp measurements are the same: in any measurement processes more complicated than a simple, non-repeated discrete measurement, reliably getting the physics right requires treating the system's behaviour unitarily, and if necessary physically modelling the measurement process. Collapse is \emph{at best} an unreliable shorthand. And of course, it is \emph{only} in ``measurement processes more complicated than a simple, non-repeated discrete measurement'' that the collapse rule could play any role in physical practice anyway. If we measure the system once and immediately discard it, the Born rule is all we need.

I conclude that the collapse postulate plays, and can play, no real part in actual applications of quantum mechanics.

\section{Against the eigenvalue-eigenvector link: problems for position}\label{vsEE-position}

Consider a mass-$m$ point particle --- either a fundamental particle, or, more typically, the centre-of-mass degree of freedom of some rigid body like a dust mote or a table. Restricting it, for simplicity, to one dimension, its most significant observables are position and momentum, corresponding to operators $\op{X}$ and $\op{P}$ respectively, obeying the commutation relation $\comm{\op{X}}{\op{P}}=i\hbar$.

It is a standard result of quantum mechanics\footnote{And, like most `standard results of quantum mechanics', there are some tacit additional mathematical assumptions required. See \cite[ch.3]{ruetschebook} for details.} that:
\begin{itemize}
\item $\op{X}$ and $\op{P}$ have continuous spectra (reflecting the fact that these are not quantised quantities, that any  position or momentum is a possible result of a measurement), and can be expressed as
\be 
\op{X}=\int_{-\infty}^{+\infty}\dr{x} x \proj{x}{x}\,\,\,\,\,\mbox{and}\,\,\,\,\,\op{P}=\hbar\int_{-\infty}^{+\infty}\dr{k} k \proj{k}{k}.
\ee
\item Any quantum state can be expressed in the position basis as
\be 
\ket{\psi}=\int_{-\infty}^{+\infty}\dr{x} \ket{x}\bk{x}{\psi}\equiv\int_{-\infty}^{+\infty}\dr{x}\psi(x) \ket{x}
\ee
where $\psi(x)\equiv \bk{x}{\psi}$ is the \emph{position-space wavefunction} (or often just \emph{wavefunction}) of the state.
\item Similarly, any quantum state can be expressed in the momentum basis as 
\be 
\ket{\psi}=\int_{-\infty}^{+\infty}\dr{k} \ket{k}\bk{k}{\psi}\equiv\int_{-\infty}^{+\infty}\dr{k}\hat{\psi}(k) \ket{k}
\ee
where $\psi(k)\equiv \bk{k}{\psi}$ is the \emph{momentum-space wavefunction} of the state.
\item The position and momentum bases are related by
\be 
\ket{k}=\frac{1}{\sqrt{2\pi}}\int_{-\infty}^{+\infty}\dr{x}\e{ikx}\ket{x}
\ee
from which it follows that the position and momentum representations are Fourier transforms of one another:
\be 
\hat{\psi}(k) = \frac{1}{\sqrt{2\pi}}\int_{-\infty}^{+\infty}\dr{x}\e{-ikx} \psi(x).
\ee
\end{itemize}
Suppose we apply the E-E link to the position of the particle. For the particle to \emph{definitely have position $x$}, it would need to be an eigenstate of $\op{X}$ --- that is, it would need to be in state $\ket{x}$. That isn't possible: because the spectrum of the position operator is continuous, the eigenstates of position are so-called `improper eigenstates' --- at least as QM is normally used, they do not represent an actually-attainable state of a quantum system. (`Legal' quantum states are normalised --- $\bk{\psi}{\psi}=1$ --- whereas the norm $\bk{x}{x}$ is infinite, or at any rate undefined.)

So: no system has a \emph{perfectly} definite position. This is not in itself problematic. It is a standard result of functional analysis that functions $f(\op{X})$ may be defined by
\be 
f(\op{X})=\int_{-\infty}^{+\infty}\dr{x} f(x) \proj{x}{x}.
\ee
In particular, if $\Sigma$ is some compact (\iec, closed and bounded) subset of the real numbers, and if $\Lambda_\Sigma$ is defined by
\begin{eqnarray}
\Lambda_\Sigma(x)& =1 &  \mbox{ if } x\in \Sigma \nonumber \\
& =0 & \mbox{otherwise}\\
\end{eqnarray}
--- that is, if $\Lambda_\Sigma$ represents the property of being in $\Sigma$ --- then
\be 
\Lambda_\Sigma(\op{X})=\int_\Sigma \proj{x}{x}
\ee
projects, according to the E-E link, onto all and only those states which are definitely located in $\Sigma$. In the position representation, this is all and only states whose wavefunction vanishes outside $\Sigma$. This suggests that if we want to represent a reasonably-well-localised particle, we should choose one with a wavefunction confined to some reasonably small $\Sigma$. (And similarly \emph{mutatis mutandis} if we want to consider systems localised in \emph{momentum}.) 

The first thing to say about this is that it is not how physicists \emph{in fact} represent localised particles. The standard strategy in physics is to represent a particle localised at some point $x_0$ by a Gaussian, \iec a state with wavefunction
\be 
\psi(x)= \mc{N}\exp(-(x-x_0)^2/2L^2).
\ee
$\psi(x)$ is very small when $|x-x_0|/L \gg 1$, so a state like this, if its position is measured, is nearly certain to be found within a few multiples of $L$ from $x_0$ --- $L$ is the ``effective width'' of the state, the size of the region in which it is `effectively localised' in physics parlance. But $\psi(x)\neq 0$ for every value of $x$ there is --- so according to the E-E link, the particle is completely delocalised, no matter how small $L$ might be.

Perhaps not too much should be made of this. Physicists use Gaussians because they are mathematically very convenient, rather than from some deep commitment to what `true' localisation is like. Perhaps we should think of the Gaussian as just a very convenient approximation to a `really' localised state, with the latter having a wavefunction with genuinely compact support. 

But suppose that a quantum system, at some initial time, \emph{does} have such a wavefunction --- say, $\psi_0$, which is localised inside some compact region $\Sigma$. If we represent that same system instead in the momentum-space representation --- that is, with its momentum-space wavefunction, which is the Fourier transform $\hat{\psi}$ of $\psi$ --- then we find that $\hat{\psi}$ does not itself have compact support. (This is a consequence of the classical Paley-Wiener theorem,\footnote{See, \egc, \citeN[pp.196-202]{rudinfunctional}.} which says \emph{inter alia} that the Fourier transform of a compactly supported $L^2$ function is holomorphic.) Via the $E-E$ link, this tells us that any particle whose \emph{position} is not completely indefinite has a completely indefinite \emph{momentum}.

That might itself be worrying: we might have hoped, given the uncertainty principle, that a particle could definitely have both (a) a position within some region of width $L$ and (b) a momentum within some region of width $\hbar/L$, but we won't get that from the $E-E$ link. But worse is to come: for consider the \emph{time evolution} of this system. We might expect that a particle whose momentum is completely indefinite will spread out instantaneously over all of space --- and indeed, this is exactly what happens. Even `confining' the system inside some potential well will not prevent its spreading out, for a quantum system will `tunnel' through any potential barrier unless it is infinite, and so unphysical.

So: no body can be localised to any degree at all for more than an instant, at least on the E-E definition of `localised'. (I have argued for this on intuitive physical grounds but will supply a mathematical proof in the next section). No assumption about the `microscopic' nature of the body in question has been made: the argument applies as readily to chairs, tables and planets as to electrons or atoms, and so chairs, tables and planets, according to the E-E link, have at almost all times a completely indefinite location. If we assume the projection postulate, of course, a system \emph{is} localised immediately after a position measurement --- but the operative word is `immediately'. An arbitrarily short time after the measurement, delocalisation is complete.

I conclude that the E-E link is of no use in understanding what it really is for a physical system to be localised to any degree.\footnote{The line of argument here has some resemblance to that used by \citeN{albertloewer1996} to argue that the E-E link should be rejected in the GRW theory in place of a ``fuzzy link''. But Albert and Loewer attributed the problem to the Gaussian collapse function used in the GRW theory, whereas as we have seen, the problem arises even in the absence of any collapse event, as a consequence of ordinary Schr\"{o}dinger dynamics.} A fortiori, it cannot be being used in physics to do useful work in our understanding of localisation. As we shall see, this is not a feature unique to spatial localisation.

\section{Against the eigenvector-eigenvalue link: problems for basically any quantity}\label{vsEE-hegerfeldt}

Hegerfeldt's theorem \cite{hegerfeldta,hegerfeldtb} is as follows:
\begin{quote}
\textbf{Hegerfeldt's theorem:} Suppose that the spectrum of the Hamiltonian of some quantum system is bounded below (something that holds of essentially any physically reasonable Hamiltonian) and let $\ket{\psi(t)}$ be some dynamical history of that system (\iec, some solution to the Schr\"{o}dinger equation). Then if $\op{\Pi}$ is any projection operator\footnote{In fact, it suffices for $\op{\Pi}$ to be a positive operator.}, \emph{exactly one} of the following holds:
\begin{enumerate}
\item $\matel{\psi(t)}{\op{\Pi}}{\psi(t)} \neq 0$ for all times $t$ except for some nowhere dense set of times of measure zero (\iec, for all times except some set of isolated instants);
\item $\matel{\psi(t)}{\op{\Pi}}{\psi(t)}=0$ for all times $t$.
\end{enumerate}
\end{quote}
Hegerfeldt proved the theorem as part of an investigation into localisation in relativistic quantum mechanics\footnote{See \citeN{halvorsonclifton} for discussion of its significance in this context.} but in fact it causes severe difficulties for the E-E link in general. For consider again our operator
\be 
\op{O}=\sum_i o_i \op{\Pi}_i
\ee
(which includes, as a special case, the sort of discretisations of position we considered previously.) The observable $O$ corresponding to \op{O} will according to the E-E link, definitely \emph{not} have value $o_i$ with respect to state $\ket{\psi(t)}$ iff $\matel{\psi(t)}{\op{\Pi}_i}{\psi(t)}=0$. So Hegerfeldt's theorem can be rephrased as 
\begin{quote}
\textbf{Hegerfeldt's theorem (indefiniteness form):} Given a system evolving unitarily over some interval of time under a Hamiltonian whose spectrum is bounded below, a given property is either (a) definitely \emph{not} possessed at every time in that interval, or (b) \emph{not} definitely not possessed at almost every time in that interval.
\end{quote}

Put another way: suppose there is some property that, at some time in the indefinitely distant future, the system might have some probability to be found to possess. Then, according to the E-E link, it is \emph{immediately} --- that is, within an arbitrarily short window of time --- indefinite whether the system has that property.

Put yet another way: anything that might at some future point be indefinite will be indefinite immediately. This seems to render the E-E link fairly useless as a description of ontology. We might have imagined that systems begin having some definite value of a given quantity, then gradually evolve so as to be indefinite across several values of that quantity, and in due course become completely indefinite with respect to that quantity (perhaps until some wavefunction collapse restores definiteness). But dynamically, that can't happen: indefiniteness is immediate if it is going to happen at all.

(I should, in fairness, acknowledge one context in which we seem to be able to get some content out of the E-E link even given Hegerfeldt's theorem. In the specific case of angular momentum (including both orbital angular momentum of some bound system, and the intrinsic spin of a particle), we could imagine the angular momentum \emph{precessing}, so that the state at time $t$ is an eigenstate of angular momentum with respect to some angle $\Omega(t)$. The system would then at all times have a definite angular momentum even though it would only be definite for an instant with respect to angular momentum in a given direction. But this relies on special features of the angular-momentum case, in particular does not generalise to position and momentum, and looks highly likely to be unstable once angular momentum couples to other degrees of freedom.)

To push the consequences of Hegerfeldt's theorem further (and also provide a rigorous justification of the claims of the previous section), suppose $\matel{\psi(t)}{\op{\Pi}}{\psi(t)}=0$ for all $t$, and consider
\be 
\mc{S}=\mbox{Span}\{\ket{\psi(t)}\}.
\ee
\mc{S} is time invariant, and so must be spanned by (possibly improper) eigenstates of the Hamiltonian. And of course any element of $\mc{S}$ is an eigenstate of $\op{\Pi}$ with eigenvalue $0$. So we can conclude that:
\begin{quote}
\textbf{Complete indefiniteness corollary:} Unless some operator has some (possibly improper) eigenstates in common with the Hamiltonian, its associated quantity is completely indefinite at almost every time.

(Readers uncomfortable with my casual use of improper eigenstates can just rephrase the requirement as ``\op{O} has an eigensubspace invariant under the Hamiltonian''.)
\end{quote}
As an application of this result, suppose that the Hamiltonian is non-degenerate (that is: has no two eigenstates with the same eigenvalue). Then a necessary and sufficient condition for a quantity not to be almost always completely indefinite is that it is a function of the Hamiltonian.

As another, consider some collection of scalar particles interacting via some potential:
\be 
\op{H}=\sum_i \frac{\op{P}_i^2}{2m_i} + V(\op{X}_1,\ldots \op{X}_n),
\ee
for some smooth function $V$. In every case I know, and in particular in the case of free particles, the eigenfunctions of this Hamiltonian have only isolated zeroes. (This is easily provable in the case where $V$ is a polynomial or other holomorphic function, so that the eigenfunctions themselves are holomorphic; it also follows in one dimension from the uniqueness theorem for solutions of ordinary differential equations.) But in that case, no eigenfunction has compact support, so no projector onto localised states is time-invariant. It follows that every particle has a completely indefinite position almost always. 

The underlying problem here is a radical mismatch between the E-E link and the way quantum mechanics \emph{actually} handles the idea of a system's becoming more spread out (speaking loosely) with respect to a given quantity. QM handles the latter through probabilities: the likelihood of a particle localised at $x$ being found very far from $x$ is initially negligibly small and only gradually increases --- and, depending on the dynamics, may never increase beyond negligible levels. But the E-E link is all-or-nothing: as soon as the system has any probability, even $10^{-10^{20}}$, of being found in some region, it is completely indefinite whether it is in that region.

I conclude that statements about a system's properties that rely on the E-E link convey essentially no information about a system between measurements (and, at the instant of measurement, everything empirically salient is coded in the Born rule in any case). As such, the E-E link cannot plausibly play a role in orthodox QM.

\section{Quantum mechanics in practice: the lab view}\label{labview}

Neither the collapse rule nor the E-E link can be part of orthodox, \iec actually-used-in-practice, QM. It isn't that they are ultimately unsatisfactory on philosophical grounds, but rather that they are not even \emph{prima facie} satisfactory, and lead to nonsense (all systems maximally indefinite all the time) or violation of empirical predictions (all measurements non-disturbing; continuous measurement impossible to define without Zeno freezing).

So what \emph{do} physicists do, if they don't do ``orthodox'' QM as it is usually understood? In the rest of the paper I shall consider two paradigms to describe orthodox QM. The first --- the ``lab view'' --- is not uncommonly found in more careful foundational discussions of QM in the physics literature (especially in quantum information) but is not ultimately satisfactory to account for physical practice; the second --- the ``decoherent view'' --- does, I think, provide an adequate fit to physical practice.

In the Lab View (my presentation here is modelled on \citeN{peres}), any application of QM should be understood as applying to some experimental setup, and that setup in turn is broken into three processes:
\begin{enumerate}
\item State preparation;
\item Dynamics;
\item State measurement.
\end{enumerate}
The first and last of these are \emph{primitive}: the question of how the system is prepared in a given state, and how it is measured, are external to the experiment and so not modelled in the physics. Only the second is regarded as a modelled physical process.

In quantum mechanics, in particular: 
\begin{enumerate}
\item The system is prepared in a state represented by some (pure or mixed) Hilbert-space state;
\item It evolves under the Schr\"{o}dinger equation for some fixed period of time;
\item The outcome of the measurement is given by the Born Rule.
\end{enumerate}
The Lab View itself does not force a unique interpretation of the underlying physics, but it is often presented in parallel with a particular interpretation, each of which is sometimes claimed as `orthodoxy'. Particularly prominent examples include:
\begin{description}
\item[Straight operationalism:] There is nothing more to quantum mechanics than a calculus that connects preparation processes (conceived of macroscopically and phenomenologically) to measurement processes (likewise conceived of); physics neither needs, nor can accommodate, any microscopic story linking the two. 

Straight operationalism is perhaps the closest realisation in mainstream physics of the old logical-positivist conception of the philosophy of science; it seems to have been more or less Heisenberg's preferred approach, and has been advocated more recently by 
Peres (\citeNP[pp.373--429]{peres},  \citeNP{fuchsperes}). The `quantum Bayesianism' or ``QBism'' of Fuchs \emph{et al} (\citeNP{fuchsinformation,fuchsmerminschack,fuchsschackgreeks}) has much in common with straight operationalism, although it holds out for some objective physical description at a deeper level (see \citeN[pp.188--235]{timpsonbook} for a critique).

\item[Complementarity:] It \emph{is} possible to describe a physical system at the microscopic level, but the appropriate description depends on the experimental context in question. Is an electron a wave or a particle? If you're carrying out a two-slit experiment, it's a wave; if you change experimental context to check which slit it went down, it's a particle. 

Niels Bohr is the most famous proponent of complementarity, though he tended to describe it in qualitative terms and engaged little with modern (Schr\"{o}dinger-Heisenberg-Dirac) quantum mechanics. \citeN{saunderscopenhagen} provides a rational reconstruction of complementarity in modern terminology; the approaches of Omnes~(\citeyearNP{omnes,omnes92,omnesbook}) and Griffiths~(\citeyearNP{griffiths,Griffiths1993,griffiths96}) are very much in the spirit of complementarity. 
\end{description}

But most relevant for our purposes is:
\begin{description}

\item[Measurement-induced collapse:] The system can be described in microscopic terms, and in a way independent of the measurement process: the physical quantities of the system are represented by the state, via the E-E link. But at the final moment of measurement at the end of the experiment, the Projection Postulate is applied, jumping the system into 
an eigenstate of the quantity being measured.
\end{description}
Here we might seem to find a rehabilitation of orthodoxy. But consider: (i) as we have seen, the E-E link in practice tells us nothing about the physical state of the system between preparation and measurement, for it is almost certain that the system is maximally indefinite with respect to any quantity of interest pretty much throughout its evolution; (ii) collapse, occurring as it does at the very end of the physical process described by the Lab View, can do no actual work in physical predictions beyond what we already get from the Born rule.

(As a terminological aside, although contemporary physics often uses ``Copenhagen interpretation'' to refer to measurement-induced collapse, the historical views developed under that name are closer to complementarity and to straight instrumentalism. See \citeN{cushingcopenhagen} and  \citeN{saunderscopenhagen} for further discussion.)

But in any case, the Lab View is itself insufficient to do justice to actual applications of QM, once they transcend the prepare-evolve-measure framework we have considered so far.

\section{Limitations of the Lab View}\label{labview-limitation}

We have already considered situations that go beyond what the Lab View, strictly speaking, can handle: those when the measurement process is not the end of our interaction with the system, where measurements are repeated or continuous.  Furthermore, and even outside these cases, the Lab View's stipulation that measurement is primitive is itself in conflict with physical practice: measurement devices are physical systems, made from atoms and designed and built on the assumption that their behaviour is governed by physical laws.

\emph{Which} laws? Back in the glory days of the Copenhagen intepretation, perhaps it was possible to suppose that the workings of lab equipment should be analysed classically, but in these days of quantum optics, superconducting supercolliders and gravity-wave-sensitive laser interferometers, we cannot avoid making extensive reference to quantum theory itself to model the workings of our apparatus. And now a regress beckons: if we can understand quantum theory only with respect to some experimental context, what is the context in which we understand the application of quantum theory to the measurement itself?

The method used is in each case the same (and we have already seen it play out in our discussion of the projection postualte):
\begin{enumerate}
\item Insofar as the physics of the measurement process are relevant, we expand the analysis to include the apparatus itself as part of the quantum system. (In quantum information this move has come to be known as `the Church of the Larger Hilbert Space'.)
\item We avoid  infinite regress by treating the Born-Rule-derived probability distribution over \emph{macroscopic} degrees of freedom not as a probability of \emph{getting certain values on measurement}, but as a probability of \emph{certain values already being possessed}.
\end{enumerate}

The need for an objective, non-quantum, macroscopically applicable language to describe the physics of measurement was already recognised by Bohr (and is acknowledged in more sophisticated operationalist accounts of quantum theory; cf. \citeN[423--427]{peres}). But it is really a special case of a more general requirement, for modern applications of quantum mechanics go beyond cases where measurement is repeated or continuous and embrace cases where we cannot really avoid interpreting the QM probabilities as entirely separate from a formal `measurement' process. 

This is particularly clear in cosmology, where it has long been suggested that Lab View quantum mechanics is unsuitable  simply because cosmology concerns the whole Universe, and so there is no `outside measurement context'; indeed, it was for exactly these reasons that Hugh Everett developed his approach to quantum theory in the first place \cite{everett}.

However, this slightly misidentifies the problem. Cosmology is concerned with the Universe on its largest scales, but not with every last feature of the Universe: realistic theories in cosmology concern particular degrees of freedom of the universe (the distribution of galaxies, for instance) and we can perfectly well treat these degrees of freedom as being measured via their interaction with other degrees of freedom outside the scope of those theories \cite{fuchsperes}. 

But there is a problem nonetheless. Namely: the processes studied in cosmology cannot, even in the loosest sense, be forced into the Lab View. They are (treated as) objective, ongoing historical processes, tested indirectly via their input into other processes; they are neither prepared in some state at the beginning, nor measured at the end, and indeed in many cases they are ongoing.

Nor is the issue specific to cosmology. The luminosity of the Sun, for instance, is determined in part via quantum mechanics: in particular, via the quantum tunneling processes that control the rate of nuclear fusion in the Sun's core as a function of its mass and composition. We can model this fairly accurately and, on the basis of that model, can deduce how the Sun's luminosity has increased over time. Astrophysicists pass that information to climate scientists, geologists, and paleontologists, who feed it into their respective models of prehistoric climates, geological processes, and ecosystems. All good science --- but only in the most Procrustean sense can we realistically regard a successful fit to data in a paleoclimate model as being a measurement of the nuclear fusion processes in the Sun a billion years ago.\footnote{This is an instance of Quine's classic objection to logical positivism~\cite{quinedogma} --- the empirical predictions of particular applications of quantum mechanics cannot be isolated from the influence of myriad other parts of our scientific world-view.}

Issues of this kind abound whenever we apply quantum theory outside stylised lab contexts. (Is the increased incidence of cancer due to Cold War nuclear-weapons tests a quantum measurement of the decay processes in the fallout products of those tests? Again, only in the most Procrustean sense.) In each case we seem to have extracted objective facts about the unobserved world from quantum theory, not merely to be dealing with a mysterious microworld that gets its meaning only when observed. But they are particularly vivid in cosmology, which is a purely observational science, and a science chiefly concerned not with repeating events in the present but with the historical evolution of the observed Universe as a whole.

As perhaps the most dramatic example available --- and probably the most important application of quantum theory in contemporary cosmology --- consider the origin of structure in the Universe. Most of that story is classical: we posit a very small amount of randomly-distributed inhomogeneity in the very-early Universe, and then plug that into our cosmological models to determine both the inhomogeneity in the cosmic microwave background and the present-day distribution of galaxies. The latter, in particular, requires very extensive computer modelling that takes into account astrophysical phenomena on a great many scales; it cannot except in the most indirect sense be regarded as a `measurement' of primordial inhomogeneity. Quantum theory comes in as a proposed source of the inhomogeneity: the posited scalar field (the `inflaton field') responsible for cosmic inflation is assumed to be in a simple quantum state in the pre-inflationary Universe (most commonly the ground state) and quantum fluctuations in that ground state, time-evolved through the inflationary era, are identified with classical inhomogeneities. Quantum-mechanical predictions thus play a role in our modelling of the Universe's history, but not a role that the Lab View seems remotely equipped to handle.

\section{Quantum mechanics in practice: the decoherent view}\label{decoherentview}

Once again: the point is not that the orthodoxy of the Lab View is conceptually inadequate, and so we must seek an unorthodox alternative; it is that  physicists manifestly \emph{are} doing quantum mechanics in regimes beyond the reach of the Lab View, so they must \emph{already} have a method for applying it that goes beyond the Lab View.

In fact, the method is fairly obvious. The probability distribution over certain degrees of freedom --- solar energy density, radiation rate, modes of the inflaton field --- is simply treated as objective, as a probability distribution over actually-existing facts, and not merely as something that is realised when an experiment is performed. So we can say, for instance, not merely that a given mode of the primordial inflaton field \emph{would have had} probability such-and-such of having a given amplitude if we were to measure it (whatever that means operationally), but that it \emph{actually did have} probability such-and-such of that amplitude.

Now, it's tempting to imagine extending this objective take on quantum probabilities to \emph{all} such probabilities: to interpret a quantum system as having some objectively-possessed value of every observable, and the quantum state as simply an economical way of coding a probability distribution over those observables. But of course, this cannot straightforwardly be done. A collection of formal results --- the Kochen-Specker theorem~(\citeNP{kochenspecker,bell1966}, \citeNP[pp.119--152]{redheadbook}, \citeNP{merminhiddenvariables}); Gleason's theorem~(\citeNP{gleason}, \citeNP[pp.27--9]{redheadbook}, \citeNP[pp.190--195]{peres} Caves \emph{et al}~\citeyearNP{cavesfuchsgleason}); the Bub-Clifton theorem~\cite{bubclifton,bubcliftongoldstein}; the PBR theorem and its relatives~\cite{puseyetal,maroney2012,Leifer2014}) --- establish that reading quantum mechanics along these lines as bearing the same relation to some underlying objective theory as classical statistical mechanics bears to classical mechanics  is pretty much\footnote{A more precise statement would be ``impossible unless that underlying objective theory has a number of extremely pathological-seeming features.'' It is not universally accepted that this rules out such theories, though; see, \egc, \citeN{spekkens} and \citeN{Leifer2014} for further discussion.} impossible.

In fact, the central problem can be appreciated without getting into the details of these results. To take an objective view of some physical quantity is to suppose that the quantity has a definite value at each instant of time, so that we can consider the various possible \emph{histories} of that quantity (that is: the various ways it can evolve over time) and assign probabilities to each. But the phenomena of interference means that this does not generically work in quantum mechanics. The quantum formalism for (say) the two slit experiment assigns a well-defined probability $P_1(x)$ to the history where the particle goes through Slit One and then hits some point $x$ on the screen, and a similarly-well-defined probability $P_2(x)$ to it hitting point $x$ via Slit Two, but of course the probability of it hitting point $x$ at all (irrespectively of which slit it goes through) is not in general $P_1(x)+P_2(x)$. So the `probabilities' assigned to these two histories do not obey the probability calculus. And things that don't obey the probability calculus are not probabilities at all.

At a fundamental level, the problem is that quantum mechanics is a dynamical theory about amplitudes, not about probabilities. The \emph{amplitudes} of the two histories in the two-slit experiment sum perfectly happily to give the amplitude of the particle reaching the slit, but amplitudes are not probabilities, and in giving rise to probabilities they can cancel out or reinforce.

However, in most physical applications of quantum theory we are \emph{not} working `at a fundamental level', which is to say that we are not attempting the usually-impossible task of deducing (far less interpreting) the evolution of the full quantum state over time. Rather, we are interested in finding higher-level, emergent dynamics, whereby we can write down dynamical equations for, and make predictions about, certain degrees of freedom of a system without having to keep track of all the remaining degrees of freedom. In the examples of the previous section, for instance, we have considered:
\begin{itemize}
\item The robust relations between macrostates of measuring devices and states of the system being measured, abstracting over the microscopic details of the measuring devices
\item The bulk thermal properties of the core of the Sun, abstracting over the vast number of microstates compatible with those bulk thermal properties
\item The low-wavelength modes of the inflaton field which are responsible for primordial inhomogeneities, abstracting over the high-wavelength degrees of freedom and the various other fields present.
\end{itemize}
In each case, we can derive from the quantum-mechanical dynamics an autonomous system of dynamical equations for these degrees of freedom. In each case, we can also derive from the Born Rule a time-dependent probability distribution over the values of those degrees of freedom. And in each case, that probability distribution defines a probability over histories that obeys the probability calculus. In each case, then, we are justified --- at least formally, if perhaps not philosophically --- in studying the autonomous dynamical system in question  as telling us how these degrees of freedom are actually evolving, quite independently of our measurement processes.

To put the position intentionally crudely: orthodox QM, I am suggesting, consists of  shifting between two different ways of understanding the quantum state according to context: interpreting quantum mechanics realistically in contexts where interference matters, and probabilistically in contexts where it does not. \emph{Obviously} this is conceptually unsatisfactory (at least on any remotely realist construal of QM) --- it is more a description of a \emph{practice} than it is a stable \emph{interpretation}. But why should that be surprising? Philosophers have spent decades complaining that physicists' approach to QM is philosophically unsatisfactory, after all. In a way, philosophers' version of `orthodoxy' does physicists too much credit in providing a self-consistent realist account of QM that just lacks a satisfactory account of exactly when collapse happens, even as it does them too little credit in failing to recognise the unsuitability of the orthodox version of orthodoxy to physical practice.

And in fact, physics has made considerable progress in clarifying just when we can, and cannot, get away with a probabilistic interpretation of the quantum state, and in particular in helping us understand why we can reliably get away with it in macroscopic contexts. The \emph{decoherence theory} developed by, \emph{inter alia}, Joos and Zeh~(\citeNP{jooszeh85,zeh93}), Zurek~(\citeyearNP{zurek91,zurekroughguide}), Gell-Mann and Hartle~\citeyear{gellmannhartle93},  Omnes~(\citeyearNP{omnes,omnes92}), and Griffiths~(\citeyearNP{griffiths,Griffiths1993}) in the 1980s and 1990s is concerned precisely with when the quantum state can be treated as probabilistic, understood either (in the environment-induced decoherence framework of Joos, Zeh and Zurek) because interference is suppressed with respect to some basis, or directly (in the consistent-histories framework of Griffiths, Omnes, and Gell-Mann and Hartle) by finding a consistent rule to assign probabilities to histories.\footnote{Appreciating that this is the task being performed by decoherence in contemporary physical practice also goes some way to explaining why the physics community has regarded decoherence as a major step towards understanding the interpretation of QM, something not generally shared by philosophers.(\citeN[p.230]{barrettbook} is typical: ``That decoherence destroys simple interference effects does not solve the measurement problem since it does not explain the determinateness of our measurement records \ldots In order to observe a single determinate record there must somewhere be a single determinate record.)}
Hence my name for this position: the ``decoherent view''.

And where is collapse in all this? Well, the condition of decoherence can be reinterpreted as a condition for when we can impose an explicit collapse rule without empirically contradicting quantum theory. But that `condition' is precisely the condition in which we can get away with treating the quantum state probabilistically, and from that perspective, ``collapse'' is just probabilistic conditionalisation. Of course, we can continue to think of the quantum state non-probabilistically, and use decoherence as a condition for when a \emph{physical} collapse process can be introduced, but now we are well outside the assessment of orthodoxy, and well along the path towards a proposed \emph{solution} of the measurement problem.

\section{Two applications of the decoherent view}

To illustrate the efficacy of the decoherent view in doing justice to physical practice, I consider two examples, from radically different sectors of physics: Stern-Gerlach type experiments, and the emergence of structure in the early Universe from primordial fluctuations of the inflaton field.

The Stern-Gerlach-type experiments I have in mind proceed as follows:
\begin{enumerate}
\item A beam of silver atoms emerges from a furnace.
\item That beam is split by a magnetic field and the beam corresponding to spin down in the $z$ direction is discarded.
\item The beam is subjected to a series of interference experiments.
\item The spin of the atoms in the beam in (say) the $x$ direction is measured by once again splitting the beam and measuring what fraction of atoms are in each beam.
\end{enumerate}
Initially, the spin degrees of freedom of a silver atom is in a mixed state
\be 
\rho_1 = \frac{1}{2}(\proj{+_z}{+_z} + \proj{-_z}{-_z}).
\ee
(The justification of this state comes from quantum statistical mechanics and lies outside the scope of this paper.) 
After the magnetic field is applied, the particles spin and position degrees of freedom become entangled, having state:
\be 
\rho_2 = \frac{1}{2}(\proj{+_z}{+_z}\otimes \proj{\varphi_+(t)}{\varphi_+(t)} + \proj{+_z}{+_z}\otimes \proj{\varphi_-(t)}{\varphi_-(t)})
\ee
where $\ket{\varphi_+(t)}$ and $\ket{\varphi_-(t)}$ are wavepacket states of negligible overlap. The mixed state cannot be used for interference experiments, so we can get away with treating it probabilistically. We now discard the $-$ part of the beam, and continue to operate only on the $+$ part; conditional on the silver atom still being in the apparatus, its spin state must be $+_z$, and so we update the state by conditionalising, to the pure state
\be 
\ket{\psi_3}=\tpk{+_z}{\varphi_+(t)}
\ee
(A more realistic treatment might allow for slight overlap of wavepackets, so that there is still some admixture of $\ket{-_z}$.)

Now we do a series of interference experiments with the system. At this point, treating it probabilistically will get us into trouble, so we avoid doing so: we continue to evolve the state unitarily and abjure probabilistic conditionalising.

Finally, we split the beam again, so it has form
\be 
\ket{\psi_4}=\alpha_+ \tpk{+_z}{\chi_+(t)} + \alpha_- \tpk{-_z}{\chi_-(t)}
\ee
(with the values of $\alpha_\pm$ depending on the details of the interference processes, and with $\ket{\chi_+(t)}$ and $\ket{\chi_-(t)}$ again having negligible overlap. We once again treat this probabilistically (since we are going to do no further interference experiments, and indeed are about to entangle the system with a macroscopically large measurement device) and interpret $|\alpha_\pm|^2$ as the probability that the particle's spin is in fact $\pm$.

As for primordial structure formation, it works as follows (here I follow \citeN[pp.470--474]{weinbergcosmology}, and excerpt a more detailed discussion in \citeN{wallacecosmology}). A quantum field theory (the inflaton field) is coupled to spacetime geometry in a perturbative fashion, and allowed to evolve in time. In the very early Universe the system is inherently quantum-mechanical and the mod-squared amplitudes of the various modes of the field cannot consistently be interpreted probabilistically.  But as the universe expands, the various mechanisms of decoherence come into play and --- still very early in the Universe's history --- we reach the point at which a probabilistic interpretation is consistent. At that point, we interpret those mod-squared amplitudes as probabilities of the actual modes of the inflaton field having various values; this determines a probability distribution over various possible inhomogeneities in the density of the early Universe, and that distribution is fed into cosmological simulations of structure formation. There is no measurement here, and no natural point for a collapse --- only a quantum state which, in due course, we can get away with treating probabilistically.

Thes examples probably strike the reader as uncomfortably opportunistic, even ad hoc. Indeed, they should so strike the reader. The ad hoc, opportunistic approach that physics takes to the interpretation of the quantum state, and the lack, in physical practice, of a clear and unequivocal understanding of the state --- \emph{this} is the quantum measurement problem, once the distractions of collapse and the E-E link are removed.

\section{Conclusion: the measurement problem from the perspective of contemporary physics}\label{conclusion}

Quantum mechanics, as actually practiced in mainstream physics, makes no use of the eigenstate-eigenvector link, nor of the collapse postulate. Its dynamics are unitary; the unitarily evolving quantum state is interpreted inchoately, as describing physical goings on in regimes where interference is important and as describing probabilities in regimes where it can be neglected. On pain of failure to account for interference, we cannot (it seems) consistently treat the state as probabilistic; on pain of failure to account for the probability rule, and more generally of failing to make contact with observation, we cannot (it seems) consistently treat the state as representational. The ``measurement problem'' from this perspective, is the task of taking this inchoate practice and showing how it can be justified given, as starting point, a well-defined physical theory --- where what counts as a ``well-defined physical theory'' will depend on one's general stance on scientific realism and the philosophy of science. Perhaps we can do so by showing how an ultimately physical superposition nonetheless appears emergently probabilistic (Everett's strategy); perhaps we can do so by showing that interference can after all be made sense of on probabilistic grounds (the quantum Bayesian strategy, and the $\psi$-epistemic one); perhaps we can do so by adding additional representational structure (the Bohmian strategy) or by changing the dynamics to introduce a stochastic element (the dynamical-collapse strategy) or by adopting  a conception of scientific theories that diverges from standard realism (the complementarity, quantum-logic, and instrumentalist strategies; perhaps the quantum-Bayesian strategy too).

From this perspective, the distinction between `pure interpretations' that leave the formalism of QM alone, and modificatory strategies that modify or supplement it, is clear. The (real) Copenhagen interpretation, quantum Bayesianism, and the Everett interpretation (whatever their strengths or weaknesses otherwise) fall into the former category, as would a (hypothethical) $\psi$-epistemic interpretation: their dynamics is unitary throughout, their formalism unsupplemented by hidden variables. Dynamical-collapse and hidden-variable theories are in the latter category, being committed to adding additional variables and/or to modifying the Schr\"{o}dinger equation.

From this perspective, too, the ``orthodox interpretation'' --- that is, the theory obtained by adding the E-E link and the projection postulate to unitary quantum mechanics, and deriving the Born rule from them --- is just one more modificatory strategy, and a strikingly implausible and unattractive one to boot. Perhaps some better attempt to solve the measurement problem will incorporate one or both, perhaps in modified and improved form --- but it is time to retire the theory that is based on them as a starting point for discussions of the measurement problem.

\section{Acknowledgements}

This paper has benefitted greatly from conversations with Simon Saunders and Chris Timpson, and from feedback when it was presented at the 2016 Michigan Foundations of Modern Physics workshop, especially from Jeff Barrett, Gordon Belot, and Antony Leggett.

\end{document}